\documentstyle[lprocl,11pt]{article}
\input epsf
\def\rmd{{\rm d}}

\def\rmO{{\rm O}}

\def\proof{\noindent{\sl Proof:}\kern0.6em}

\def\frac#1#2{\hbox{$#1\over#2$}}
\def\dual{\mathstrut^*\kern-0.1em}

\def\lvec#1{\setbox0=\hbox{$#1$}
    \setbox1=\hbox{$\scriptstyle\leftarrow$}
    #1\kern-\wd0\smash{
    \raise\ht0\hbox{$\raise1pt\hbox{$\scriptstyle\leftarrow$}$}}
    \kern-\wd1\kern\wd0}
\def\rvec#1{\setbox0=\hbox{$#1$}
    \setbox1=\hbox{$\scriptstyle\rightarrow$}
    #1\kern-\wd0\smash{
    \raise\ht0\hbox{$\raise1pt\hbox{$\scriptstyle\rightarrow$}$}}
    \kern-\wd1\kern\wd0}

\def\nabstar#1{\nabla\kern-0.5pt\smash{\raise 4.5pt\hbox{$\ast$}}
               \kern-4.5pt_{#1}}

\def\drvstar#1{\partial\kern-0.5pt\smash{\raise 4.5pt\hbox{$\ast$}}
               \kern-5.0pt_{#1}}

\def\MeV{{\rm MeV}}

\def\psibar{\overline{\psi}}

\def\rhoprime{\rho\kern1pt'}
\def\rhobar{\bar{\rho}}
\def\rhobarprime{\rhobar\kern1pt'}
\def\rhobartilde{\kern2pt\tilde{\kern-2pt\rhobar}}
\def\rhobartildeprime{\kern2pt\tilde{\kern-2pt\rhobar}\kern1pt'}

\def\zetabar{\bar{\zeta}}
\def\zetaprime{\zeta\kern1pt'}
\def\zetabarprime{\zetabar\kern1pt'}
\def\zetar{\zeta_{\raise-1pt\hbox{\sixrm R}}}
\def\zetabarr{\zetabar_{\raise-1pt\hbox{\sixrm R}}}

\def\phiimpr{\phi_{\kern0.5pt\hbox{\sixrm I}}}

\def\dirac#1{\gamma_{#1}}
\def\diracstar#1#2{
    \setbox0=\hbox{$\gamma$}\setbox1=\hbox{$\gamma_{#1}$}
    \gamma_{#1}\kern-\wd1\kern\wd0
    \smash{\raise4.5pt\hbox{$\scriptstyle#2$}}}

\def\csw{c_{\rm sw}}

\def\Seff{S_{\rm eff}}

\def\opprime#1{\setbox0=\hbox{${\cal O}$}\setbox1=\hbox{${\cal O}_{\rm #1}$}
    {\cal O}_{\rm #1}\kern-\wd1\kern\wd0
    \smash{\raise4.5pt\hbox{\kern1pt$\scriptstyle\prime$}}\kern1pt}

\def\ophatprime#1{\setbox0=\hbox{$\widehat{\cal O}$}
    \setbox1=\hbox{$\widehat{\cal O}_{\rm #1}$}
    \widehat{\cal O}_{\rm #1}\kern-\wd1\kern\wd0
    \smash{\raise4.5pt\hbox{\kern1pt$\scriptstyle\prime$}}\kern1pt}

\def\bopprime#1{\setbox0=\hbox{${\cal O}$}\setbox1=\hbox{${\cal O}_{\rm #1}$}
    {\cal L}_{\rm #1}\kern-\wd1\kern\wd0
    \smash{\raise4.5pt\hbox{\kern1pt$\scriptstyle\prime$}}\kern1pt}

\def\blagprime#1{\setbox0=\hbox{${\cal B}$}\setbox1=\hbox{${\cal B}_{#1}$}
    {\cal B}_{#1}\kern-\wd1\kern\wd0
    \smash{\raise5.2pt\hbox{\kern1pt$\scriptstyle\prime$}}\kern1pt}

\def\gmsbar{\bar{g}_{\kern0.5pt\smallmsbar}}
\def\gbar{\bar{g}}

\def\mbar{\overline{m\kern-1pt}\kern1pt}

\def\zp{Z_{\rm P}}
\def\xp{X_{\rm P}}
\def\zpmom{\zp^{\raise1pt\hbox{\sixrm MOM}}}
\def\ztwomom{Z_2^{\raise1pt\hbox{\sixrm MOM}}}

\def\msbar{{\rm \overline{MS\kern-0.05em}\kern0.05em}}
\def\smallmsbar{\overline{\hbox{\sixrm MS\kern-0.10em}}
                \hbox{\sixrm\kern0.10em}}
\def\lat{{\rm lat}}

\def\mpi{m_{\pi}}

\def\fK{f_{\hbox{\sixrm K}}}
\def\fps{f_{\kern-1pt\hbox{\sixrm PS}}}
\def\fpi{f_{\pi}}
\def\fv{f_{\kern-1pt\hbox{\sixrm V}}}

\def\mK{m_{\hbox{\sixrm K}}}
\def\mKplus{m_{\kern-1pt\hbox{\sixrm K}^{+}}}
\def\mKnull{m_{\kern-1pt\hbox{\sixrm K}^{0}}}
\def\mKstar{m_{\hbox{\sixrm K}^{\ast}}}
\def\mps{m_{\kern-1pt\hbox{\sixrm PS}}}
\def\mv{m_{\hbox{\sixrm V}}}

\begin{document}

\rightline{DESY 97-215}
\rightline{November 1997}
\vspace{0.5cm}

\title{THEORETICAL ADVANCES IN LATTICE QCD\,%
\footnote{Talk given at the 18th International Symposium on
Lepton-Photon Interactions, Hamburg, \\28 July -- 1 August 1997}}

\author{M.~L\"USCHER}

\address{Deutsches Elektronen-Synchrotron DESY \\
         Notkestrasse 85, D-22603 Hamburg, Germany \\
         E-mail: luscher@mail.desy.de}

\maketitle\abstracts{
The past few years have seen many interesting theoretical
developments in lattice QCD.
This talk (which is intended for non-experts)
focuses on the problem of non-perturbative renormalization and the 
question of how precisely the continuum limit is reached.
Progress in these areas is crucial in order to be able to compute
quantities of phenomenological interest,
such as the hadron spectrum, 
the running quark masses and weak transition matrix elements,
with controlled systematic errors.
}

\section{Introduction}

This year's lattice conference 
attracted more than 300 physicists from all over the world.
Lattice QCD remains to be the dominant subject at these conferences,
but other topics are also being addressed, such
as the electroweak phase transition, quantum gravity and
supersymmetric Yang-Mills theories.
Much of the work done in QCD is spent to improve 
the computations of 
hadron masses, decay constants and weak transition matrix elements. 
Calculations of moments of structure functions, 
the running coupling and quark masses and many other
physical quantities have also been reported.
A good place to look for specific results are
the proceedings of the 1996 lattice conference~\cite{LatProceedings}
and more recent contributions can be found 
in the hep-lat section of the Los Alamos preprint server.

\subsection{Numerical simulations}

Quantitative results in lattice QCD are almost exclusively obtained
using numerical simulations. Such calculations proceed by 
choosing a finite lattice, with spacing $a$ and linear extent $L$,
which is sufficiently small that the quark and gluon fields
can be stored in the memory of a computer.
Through a Monte Carlo algorithm one then generates a representative
ensemble of fields for the Feynman path integral and extracts the
physical quantities from ensemble averages. 
Apart from statistical errors this method yields exact results
for the chosen lattice parameters and is hence suitable for 
non-perturbative studies of QCD.

Numerical simulations require powerful computers and
a continuous effort for algorithm and software development.
Technical expertise is also needed to be able to cope
with the systematic errors incurred by the finiteness
of the lattice and by the data analysis.
In practice such calculations are being performed by 
collaborations of (say) 5--15 physicists.
For these groups to remain competitive it is vital 
that they have access to dedicated computer systems.
An adequate amount of computing power
would otherwise be difficult to obtain over a longer period of time.

\subsection{Computers}

Until recently leading edge numerical simulations of lattice
QCD have been performed on computers with sustained computational
speeds on the order of 10 Gflops.
The community is now moving to the next generation computers which
deliver several 100 Gflops for QCD programs.

One of these machines, the CP-PACS computer~\cite{CPPACS},
has been installed last year at the Center for 
Computational Physics in Tsukuba.
With its 2048 pro\-ces\-sing nodes, a total memory of 128 GB
and a theoretical peak speed of 614 Gflops, this computer 
is a unique research facility for lattice QCD.

Other machines that will be available for 
dedicated use by the lattice theorists 
include the QCDSP and the APEmille 
computers~\cite{QCDSP,APEmille}. 
The first of these has been designed by a consortium
of physicists in the US.
Machines of various size are being assembled and 
will be installed in the course of this year at different places,
totalling more than 1000 Gflops of peak computational power. 
The APEmille grew out of
a long-term effort of INFN, now also supported by
DESY-Zeuthen, to construct affordable computers
optimised for lattice gauge theory applications.
It has a scalable massively parallel architecture with
the largest system delivering more than 1000 Gflops.
A fully operational, medium-size machine is expected to be 
available next summer. 

It should be said at this point that
the theoretical peak speed of a computer is a relatively
crude measure of its performance.
The sustained computational speed that 
can be attained also depends on many other parameters 
such as the bandwidths for memory-to-processor and 
node-to-node communications, for example.
All computers mentioned here are well balanced in this respect
and achieve a high efficiency for QCD programs.

\subsection{Lattice QCD at 100 Gflops}

At present most calculations in lattice QCD neglect sea quark effects,
because the known simulation algorithms
slow down dramatically when they are included.
If one is willing to make this approximation
(which is called ``quenched QCD"),
the new computers are good for lattice sizes up to about $128\times64^3$. 
Such a lattice may be arranged to have a spacing
$a=0.05$ fm, for example, in which case its spatial extent will be $3.2$ fm.
This is a very comfortable situation for 
calculations of the light hadron masses and many other quantities
of interest. In general the increased computer power 
allows one to explore a greater range of lattices
with higher statistics and thus to achieve better
control on the systematic errors.

When a doublet of sea quarks is included in the 
simulations, lattice sizes up to $64\times32^3$ are expected to be
within reach. This will be quite exciting, because
studies of full QCD on large lattices have
been rare so far, leaving many basic questions unanswered.
The physics programme is essentially the same as in quenched QCD,
but since one cannot afford to perform simulations at very many
different values of the parameters, and since the generated ensembles
of field configurations tend to be smaller, 
the results will be generally less precise.


\subsection{Topics covered in this talk}

While the progress in computer technology is impressive,
one cannot ignore the fact that the accessible lattices 
are too small to accommodate very large scale differences.
This talk addresses two important issues which arise
from this limitation and which must be resolved if 
one is interested in results with reliable error bounds.

One of the problems is that the lattice spacing
cannot be made arbitrarily small compared to the 
relevant physical scales (the confinement radius for example).
Taking the continuum limit thus is a non-trivial task
and a lot of work has recently been spent
to answer the question of how precisely the limit is approached 
and whether the lattice effects are negligible 
at current values of the lattice spacing. 

The other topic that will be discussed 
goes under the heading of non-perturbative renormalization.
In physical terms the problem is to 
establish the relation between the low-energy properties of QCD
and the perturbative regime.
Hadronic matrix elements of operators,
whose normalization is specified at high energies through 
the $\msbar$ scheme of dimensional regularization,
are an obvious case where this is required.
Again a large scale difference is involved which makes a direct
approach difficult, but promising ways to solve 
the problem have now been found.

\section{Lattice effects and the approach to the continuum limit}

\subsection{Perturbation theory}

In lattice QCD one is primarily interested in the non-perturbative
aspects of the theory. 
Perturbation theory can, however, give important structural insights
and it has proved useful to study the nature of the continuum limit
in this framework. A remarkable result in this connection is
that the existence of the limit has been rigorously established
to all orders of the expansion~\cite{Reisz}. 

The Feynman rules on the lattice are derived straightforwardly
from the chosen lattice action. 
Compared to the usual rules, an important difference is
that the propagators and vertices are relatively complicated functions
of the momenta and of the lattice spacing $a$.
In particular, at tree-level all the lattice spacing dependence
arises in this way.

A simple example illustrating this is the quark-gluon vertex.
Using the standard formu\-lation 
of lattice QCD (which goes back to Wilson's famous paper 
of 1974~\cite{Wilson}), one finds

\vspace{-0.3cm plus 0.05cm}
\rightline{
\begin{minipage}{6cm}
\epsfxsize=1.5cm\vspace{0.4cm}\hspace{4.5cm}\epsfbox{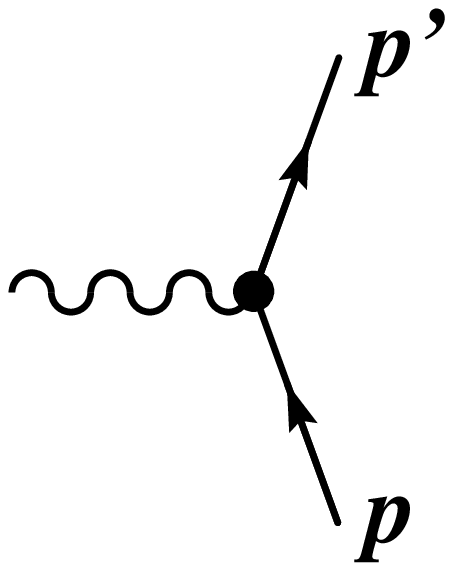}
\end{minipage}
\begin{minipage}{10cm}
\begin{equation}
  =\;g_0\lambda^a\left\{
   \dirac{\mu}+\frac{i}{2}a(p+p')_{\mu}+\rmO(a^2)\right\},
   \hspace{2.5cm}
\end{equation}
\end{minipage}}

\vspace{0.2cm plus 0.05cm}
\noindent
where $g_0$ denotes the bare gauge coupling
and $\lambda^a$ a colour matrix.
It is immediately clear from this expression that
the leading lattice corrections to the continuum term
can be quite large even if the quark momenta $p$ and $p'$ are 
well below the momentum cutoff $\pi/a$. 
Moreover the corrections violate chiral symmetry,
a fact that has long been a source of concern
since many properties of low-energy QCD depend on this symmetry.

Lattice Feynman diagrams with $l$ loops
and engineering dimension $\omega$ can be expanded in 
an asymptotic series 
of the form~\cite{SymanzikI,SymanzikII}
\begin{equation}
  a^{-\omega}\sum_{k=0}^{\infty}\sum_{j=0}^l c_{kj}a^k\left[\ln(a)\right]^j.
\end{equation}
After renormalization the negative powers in the lattice spacing
and the logarithmically divergent terms cancel in the sum of all diagrams.
The leading lattice corrections thus
vanish proportionally to $a$ (up to logarithms)
at any order of perturbation theory.

\begin{figure}[t]
\vspace{-2.3cm}
\hbox{\epsfxsize=11.0cm\hspace{2.0cm}\epsfbox{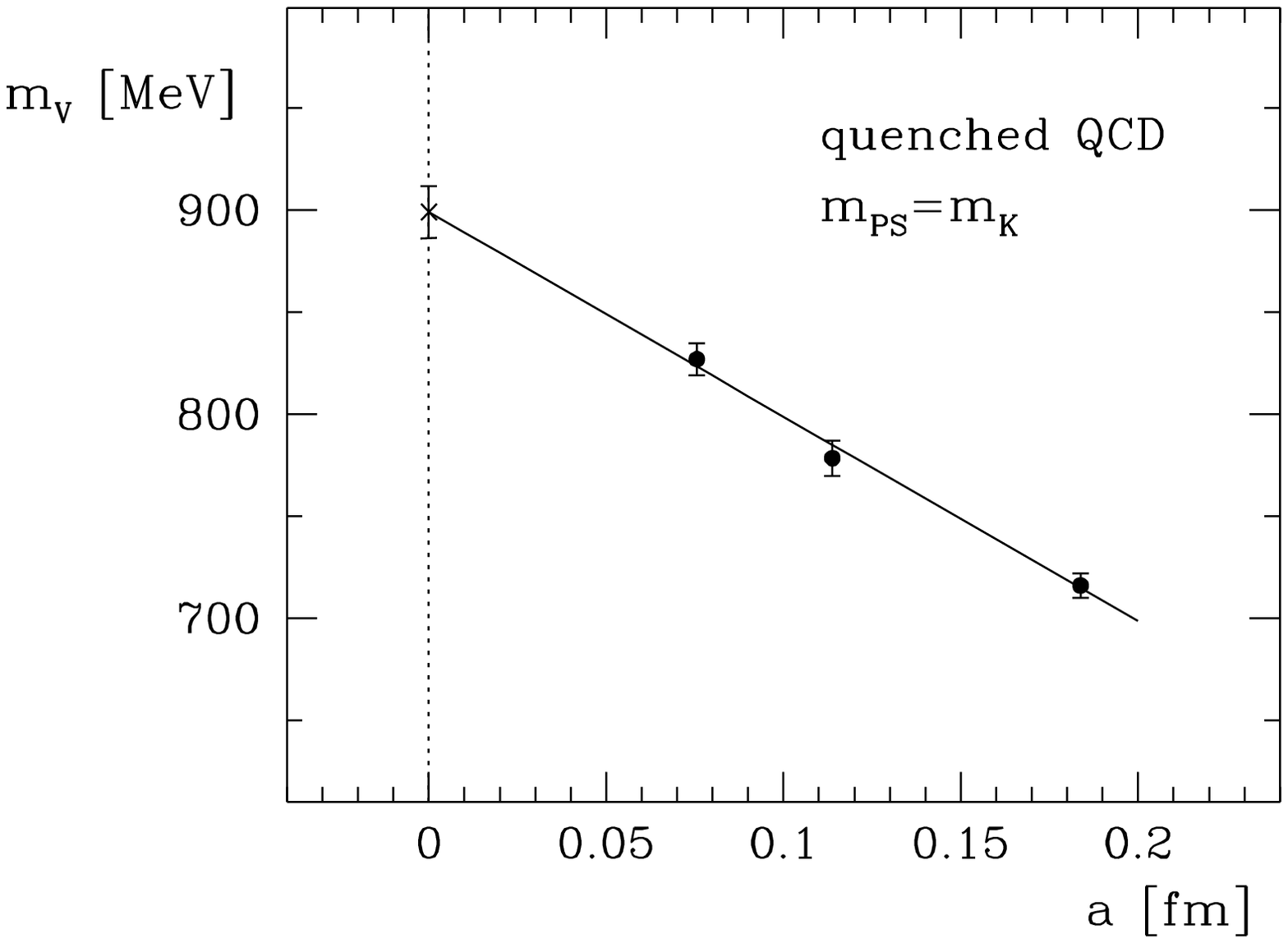}}
\vspace{-1.2cm}
\begin{center}
\footnotesize
Figure~1: Calculated values of 
the vector meson mass (full circles) and
linear extrapolation to the continuum limit (cross).
Simulation data from Butler et al.~(GF11 collab.)~\cite{WeingartenI}.
\end{center}
\vspace{0.0cm plus 0.3cm}
\end{figure}

\subsection{Cutoff dependence of hadron masses}

Sizeable lattice effects are also observed 
at the non-perturbative level when calculating hadron masses, for example.
An impressive demonstration of this is obtained as follows.
Let us consider QCD 
with a doublet of quarks of equal mass, adjusted so that
the mass of the pseudo-scalar mesons coincides with
the physical kaon mass. This sets the quark mass to about half
the strange quark mass and one thus expects
that the mass of the lightest vector meson is given by
\begin{equation}
  \mv\simeq\mKstar=892\,\MeV.
\end{equation}
Computations of the meson masses using numerical simulations
however show that this is not the case at 
the accessible lattice spacings (see Figure~1).
Instead one observes a strong dependence on the lattice spacing
and it is only after extrapolating the data to $a=0$ that one ends up
with a value close to expectations.

\subsection{Effective continuum theory}

In phenomenology it is well known that the effects of as yet
undetected substructures~or heavy particles may be described by adding
higher-dimensional interaction terms to the Stan\-dard Model lagrangian.
From the point of view of an underlying more complete theory, the
Standard Model together with the added terms then is a low-energy
effective theory. A similar situation occurs in lattice QCD, where
the momentum cutoff may be regarded (in a purely mathematical sense)
as a scale of new physics. The associated low-energy effective theory
is a continuum theory with action~\cite{SymanzikI,SymanzikII}
\begin{equation}
  \Seff=\int\rmd^4x \left\{
  {\cal L}_0(x)+a{\cal L}_1(x)+a^2{\cal L}_2(x)+\ldots\right\},
\end{equation}
where ${\cal L}_0$ denotes the continuum QCD lagrangian
and the ${\cal L}_k$'s, $k\geq1$, are linear combinations of 
local operators of dimension $4+k$ with coefficients 
that are slowly varying functions of $a$ 
(powers of logarithms in perturbation theory).

Through the effective continuum theory, 
the lattice spacing dependence is made explicit and 
a better understanding of the approach to the continuum limit is achieved.
In particular, 
neglecting terms that do not contribute to on-shell quantities, 
or which amount to renormalizations of the coupling and the quark masses,
the general expression for the leading lattice correction is
\begin{equation}
  {\cal L}_1=c_1\,\psibar\,\sigma_{\mu\nu}F_{\mu\nu}\psi,
\end{equation}
with $F_{\mu\nu}$ being the gluon field strength and $\psi$ the quark field.
The lattice thus assigns an anomalous
colour-magnetic moment of order $a$ to the quarks.
Very many more terms contribute to ${\cal L}_2$ 
and a simple physical interpretation is not easily given.
The pattern of the lattice effects of order $a^2$ 
should hence be expected to be rather complicated.

\subsection{O($a$) improvement}

The effective action, Eq.~(4), depends on the physics at the scale of the 
cutoff, i.e.~on how precisely the lattice theory is set up.
By choosing an improved lattice action one may hence be
able to reduce the size of the correction terms and thus
to accelerate the convergence to 
the continuum limit~\cite{SymanzikIII}.
Different ways to implement this idea are being explored
and there is currently no single preferred way to proceed.
At last year's lattice conference the subject has been reviewed
by Niedermayer~\cite{ImpReview} and many interesting
contributions have been made since then.

O($a$) improvement is a relatively modest approach, where 
the leading correction ${\cal L}_1$ is cancelled
by replacing the Wilson action through~\cite{SW}
\begin{equation}
  S_{\rm Wilson}+a^5\sum_{x}\csw\,
  \psibar(x)\frac{i}{4}\sigma_{\mu\nu}F_{\mu\nu}(x)\psi(x)
\end{equation}
and tuning the coefficient $\csw$. 
At the time when Sheikholeslami and Wohlert
published their paper~\cite{SW},
the proposition did not receive too much attention,
because systematic studies of lattice effects were not feasible
with the available computers. 
The situation has now changed and there is general agreement
that improvement is useful or even necessary, particularly
in full QCD where simulations are
exceedingly expensive in terms of computer time.

An obvious technical difficulty is that $\csw$
(which is a function of the bare gauge coupling) needs to be
determined accurately. The problem has only recently been solved
by studying the axial current conservation 
on the lattice~\cite{AlphaI}.
Chiral symmetry is not preserved by the lattice regularization
and the PCAC relation
satisfied by the $\bar{u}d$ component of the axial current,
\begin{equation}
  \partial_{\mu}(\bar{u}\dirac{\mu}\dirac{5}d)=
  (\mbar_{\rm u}+\mbar_{\rm d})\bar{u}\dirac{5}d+\epsilon(a),
\end{equation}
thus includes a non-zero error term.
In general $\epsilon(a)$ vanishes proportionally to $a$,
but after improvement the error is reduced to order $a^2$
if $\csw$ has the proper value. 
Conversely this may be taken as a condition fixing $\csw$,
i.e.~the coefficient can be computed by minimizing the error
term in various matrix elements of Eq.~(7).
Proceeding along these lines one has been able to calculate
$\csw$ non-peturbatively in
quenched QCD~\cite{AlphaI,SCRIa} and now also 
in full QCD with a doublet of massless sea quarks~\cite{AlphaII}.

\subsection{Impact of O($a$) improvement on physical quantities} 

Once the improvement programme has been properly implemented,
the question arises whether the lattice effects
on the quantities of physical interest 
are significantly reduced at the accessible lattice spacings.
Several collaborations have set out to check 
this~\cite{QCDSFa,UKQCDa,APETOVa},
but it is too early to draw definite conclusions.
The status of these studies
has recently been summarized by Wittig~\cite{WittigI}.

For illustration let us again consider
the calculation of the vector meson mass $\mv$ discussed in Subsection~2.2. 
A preliminary analysis of simulation results from
the UKQCD collaboration gives, for the O($a$) improved theory,
$\mv=924(17)$ MeV at $a=0.098$ fm and $\mv=932(26)$ MeV at $a=0.072$ fm.
These numbers do not show any significant 
dependence on the lattice spacing and they are also compatible with 
the value $\mv=899(13)$ that one obtains through extrapolation
to $a=0$ of the results from the unimproved theory (left-most point in
Figure~1). 

Other quantities that are being studied include the pseudo-scalar
and vector meson decay constants and the renormalized quark masses.
The experience accumulated so far suggests that 
the residual lattice effects are indeed small if $a\leq0.1$ fm.
Most experts would however agree that further confirmation is still
needed.

\subsection{Synthesis}

At sufficiently small lattice spacings 
the effective continuum theory provides an elegant description of 
the approach to the continuum limit.
Whether the currently accessible lattice spacings
are in the range where the effective theory applies
is not immediately clear,
but the observed pattern of the lattice spacing dependence in the 
unimproved theory and the fact that O($a$) improvement 
appears to work out strongly indicate this to be so.
Very much smaller lattice spacings are then not required
to reliably reach the continuum limit. 
It is evidently of great importance to put this conclusion
on firmer grounds by continuing and extending the ongoing 
studies of O($a$) improvement and other forms of improvement.

\vfill\eject

\section{Non-perturbative renormalization}

\subsection{Example}

We now turn to the second subject covered in this talk
and begin by describing one of the standard ways to compute 
the running quark masses in lattice QCD.
The need for non-perturbative renormalization will then become clear.
Any details not connected with this particular aspect of 
the calculation are omitted.

A possible starting point to obtain the sum 
$\mbar_{\rm u}+\mbar_{\rm s}$ of the up and 
the strange quark masses is the PCAC relation
\begin{equation}
  \mK^2f_{\lower1.0pt\hbox{\sixrm K}}=(\mbar_{\rm u}+\mbar_{\rm s})
  \langle0|\,\bar{u}\dirac{5}s\,|K^{+}\rangle.
\end{equation}
Since the kaon mass $\mK$ and the decay constant $\fK$ are 
known from experiment, it suffices to evaluate the matrix element
on the right-hand side of this equation.
On the lattice one first computes the matrix element of 
the bare operator $(\bar{u}\dirac{5}s)_{\lat}$
and then multiplies the result with the renormalization factor $\zp$
relating $(\bar{u}\dirac{5}s)_{\lat}$ to the renormalized
density $\bar{u}\dirac{5}s$.

\renewcommand{\arraystretch}{1.3}
\begin{table}[b]
\caption{Recent results for $\mbar_{\rm s}$
(quenched QCD, $\msbar$ scheme at $\mu=2$ GeV)}
\vspace{0.2cm}
\begin{center}
\footnotesize
\begin{tabular}{|l|l|}
\hline
\multicolumn{1}{|c|}{$\mbar_{\rm s}$ [MeV]}   
& \multicolumn{1}{c|}{reference} \\
\hline
\hspace{0.5cm}$122(20)$\hspace{0.5cm}  
& Allton et al.~(APE collab.)~\cite{QM_AlltonEtAl} \\
\hspace{0.5cm}$112(5)$                      
& G\"ockeler et al.~(QCDSF collab.)~\cite{QCDSFa} \\
\hspace{0.5cm}$111(4)$                      
& Aoki et al.~(CP-PACS collab.)~\cite{QM_CPPACS} \\
\hspace{0.5cm}$\phantom{0}95(16)$           
& Gough et al.~\cite{QM_GoughEtAl}  \\
\hspace{0.5cm}$\phantom{0}88(10)$           
& Gupta \& Bhattacharya~\cite{QM_Gupta} \\
\hline
\end{tabular}
\end{center}
\end{table}
\renewcommand{\arraystretch}{1.0}

Some recent results for the strange quark mass 
obtained in this way or in similar ways are listed in Table~1
(further results can be found
in the review of Bhattacharya and Gupta~\cite{QM_Review}).
The sizeable differences among these numbers
have many causes.
An important uncertainty arises from the fact that, 
in one form or another, the one-loop formula
\begin{equation}
  \zp=1+{g_0^2\over4\pi}\left\{(2/\pi)\ln(a\mu)+C\right\}
      +\rmO(g_0^4)
\end{equation}
has been used to compute the renormalization factor,
where $g_0$ denotes the bare lattice coupling, $\mu$ the
normalization mass in the $\msbar$ scheme and $C$ a calculable constant
that depends on the details of the lattice regularization.
Bare perturbation theory has long been known to be unreliable
and various recipes, based on mean-field theory or resummations
of tadpole diagrams, have been given to deal 
with this problem~\cite{Parisi,Lepenzie}.
Different prescriptions however give different results and
it is in any case unclear how the error on the so calculated values
of $\zp$ can be reliably assessed.

\subsection{Intermediate renormalization}

An interesting method to compute renormalization factors
that does not rely on bare perturbation theory has been
proposed by Martinelli et al.~\cite{IR_Martinelli}.
The idea is to proceed in~two steps, first matching
the lattice with an intermediate momentum subtraction (MOM) scheme 
and then passing to the $\msbar$ scheme.
The details of the intermediate MOM scheme 
do not influence the final results and are of only practical
importance. One usually chooses the Landau gauge and imposes
normalization conditions on the propagators and the 
vertex functions at some momentum~$p$.
In the case of the pseudo-scalar density, for example,
the renormalization constant $\zpmom$ is defined through

\vbox{
\vspace{1.0cm}
\begin{equation}
=\;\ztwomom/\zpmom\;\times
\end{equation}}
\vbox{
\vspace{-1.35cm}
\hbox{\epsfxsize=6.0cm\hspace{4.4cm}\epsfbox{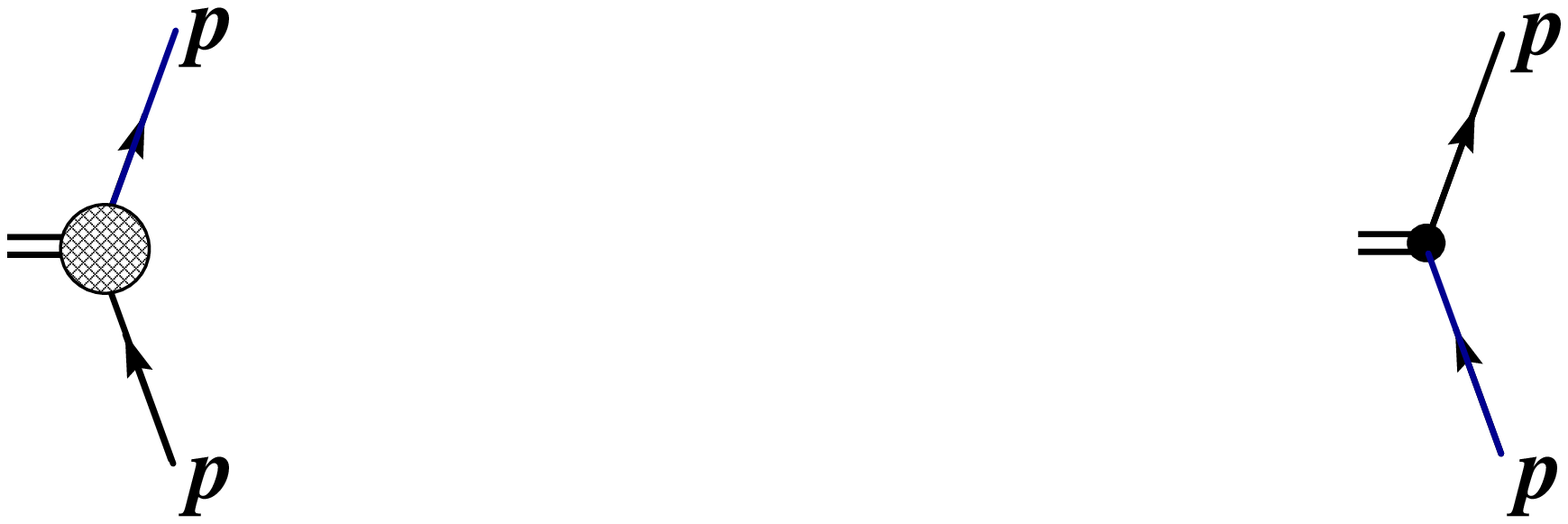}}
\vspace{0.3cm}}

\noindent
where $\ztwomom$ denotes the quark wave function renormalization constant
and the diagrams represent the full and the bare vertex function 
associated with this operator. 

On a given lattice and for a range of momenta,
the quark propagator and the full vertex function
can be computed using numerical simulations~\cite{IR_Martinelli,IR_Giusti}.
$\ztwomom$ and $\zpmom$ are thus obtained non-perturbatively.
The total renormalization factor relating the lattice normali\-zations with 
the $\msbar$ scheme is then given by
\begin{equation}
   \zp(g_0,a\mu)=\zpmom(g_0,ap)\xp(\gmsbar,p/\mu),
\end{equation}
with $\xp$ being the finite renormalization constant required to match
the MOM with the $\msbar$ scheme. $\xp$ is known
to one-loop order of renormalized perturbation theory 
and could easily be worked out to two loops.

While this method avoids the use of bare perturbation theory,
it has its own problems, the most important being 
that the momentum $p$ should be significantly
smaller than $1/a$ to suppress the lattice effects,
but not too small as otherwise one may not be confident to 
apply renormalized perturbation theory to compute $\xp$.
On the current lattices the values of $1/a$ are between
$2$ and $4$ GeV and it is hence not totally obvious that 
a range of momenta exists where both conditions 
are approximately satisfied~\cite{IR_Crisafulli,IR_Goeckeler}.
A simple criterion which may be applied in this connection
is that the calculated values of $\zp$ should be independent of~$p$.

\subsection{Non-perturbative renormalization group}

It should now be quite clear that further progress
depends on whether one is able 
to make contact with the high-energy
regime of the theory in a controlled manner. 
As has been noted some time ago, this can be achieved through a
recursive procedure~\cite{FSTa}.
A general solution of the non-perturbative
renormalization problem
is then obtained~\cite{FSTb,FSTc,FSTd,FSTe}.

The basic idea of the method can be explained in a few lines.
One begins by \hbox{introducing} a special intermediate renormalization scheme, 
where all normalization conditions are imposed at scale $\mu=1/L$
and zero quark masses, $L$ being the spatial extent of 
the lattice. We could choose a MOM scheme, for example,
and set $p$ equal to $2\pi/L$, the smallest non-zero 
momentum available in finite volume.
But this is not the only possibility and other schemes are
in fact preferred for technical reasons.

In such a scheme the scale evolution 
of the renormalized parameters and operators can be studied
simply by changing the lattice size $L$ at fixed bare parameters.
One usually simulates pairs of lattices with sizes $L$ and $2L$.
Up to lattice effects the running couplings on the two
lattices are then related through
\begin{equation}
  \gbar^2(2L)=\sigma(\gbar^2(L)),
\end{equation}
where $\sigma$ is an integrated form
of the Callan-Symanzik $\beta$-function.
Similar scaling functions are associated with 
the renormalized quark masses and the local operators.
An important point to note is that these functions can be computed 
for a large range of $\gbar^2$ without running into
uncontrolled lattice effects, because the 
lattice spacing is always much smaller than $L$,  
on any reasonable lattice, no matter how small $L$ is in physical units.

Once the scaling functions are known, one can 
move up and down the energy scale by factors of $2$.
With only a few steps a much larger range of scales can be 
covered in this way than would otherwise be possible.

\begin{figure}[t]
\begin{center}
\small
\renewcommand{\arraystretch}{2.0}
\begin{tabular}{l c l}
$\Lambda_{\smallmsbar}=k\Lambda$, $M$ 
&\hspace{0.0cm}$\longleftarrow$\hspace{1.0cm} 
&\hspace{0.5cm}$\Lambda$, $M$ \\
\hspace{1.0cm}$\downarrow$ 
& 
&\hspace{1.0cm}$\uparrow$ \\
\begin{minipage}{3.0cm}
perturbative\\
evolution
\end{minipage} 
&&
\begin{minipage}{3.0cm}
perturbative\\
evolution
\end{minipage}\\
\hspace{1.0cm}$\downarrow$ 
& 
&\hspace{1.0cm}$\uparrow$ \\
$\alpha_{\smallmsbar}(\mu)$, $\mbar_{\smallmsbar}(\mu)$ 
&&
$\gbar$, $\mbar$ at $\mu=100$ GeV \\
&
&\hspace{1.0cm}$\uparrow$ \\
&&
\begin{minipage}{3.0cm}
non-perturbative\\
evolution
\end{minipage}\\
&
&\hspace{1.0cm}$\uparrow$ \\
\begin{minipage}{3.0cm}
$\fpi,\mpi,\mK,\ldots$
\end{minipage} 
&\hspace{0.0cm}$\longrightarrow$\hspace{1.0cm}
&
\begin{minipage}{4.0cm}
finite-volume scheme\\
$\gbar$, $\mbar$ at $\mu=0.6$ GeV
\end{minipage}\\
\end{tabular}
\renewcommand{\arraystretch}{1.0}
\end{center}
\vspace{0.6cm}
\begin{center}
\footnotesize
Figure~2: Strategy to compute the running coupling and quark masses,
taking low-energy data as input and using the non-perturbative
renormalization group to scale up to high energies.
\end{center}
\vspace{0.0cm plus 0.3cm}
\end{figure}

\subsection{Application}

So far the recursive procedure described above
has been used to compute the running coupling 
in quenched QCD~\cite{FSTb,FSTc}
and first results are now also being obtained for  
the running quark masses~\cite{FSTd,FSTe}.
The calculation follows the arrows in the diagram shown in 
Figure~2, starting at the lower-left corner.
In this plot the energy is increasing from the bottom to the top
while the entries in the left and right columns refer
to infinite and finite volume quantities respectively.

The computation begins by calculating 
the renormalized coupling $\gbar$ and 
quark masses $\mbar$ in the chosen finite-volume scheme
at some low value of $\mu$,
where contact with the hadronic scales can easily be made using
numerical simulations.
In the next step one takes
these results as initial values for the non-perturbative renormalization group
and scales the coupling and quark masses to high energies.
At still higher energies the 
perturbative evolution equations apply and 
the $\Lambda$-parameter and the renormalization group invariant
quark masses 
\begin{equation}
  M=\lim_{\mu\to\infty}
  \mbar\left(2b_0\gbar^2\right)^{-d_0/2b_0}
\end{equation}
may be extracted with negligible systematic error
($b_0$ and $d_0$ denote the one-loop coefficients of the 
$\beta$-function and the anomalous mass dimension).
The renormalization group invariant quark masses $M$ 
are scheme-independent and thus do not change
when we pass from the finite-volume to the $\msbar$ scheme,
while the matching of the $\Lambda$-parameters 
involves an exactly calculable proportionality constant $k$
(top line of Figure~2). 
The perturbative evolution in the $\msbar$ scheme, which is now
known through four loops~\cite{RitbergenI,Chetyrkin,RitbergenII}, 
finally yields the running coupling and quark masses in this scheme.

\begin{figure}[t]
\vspace{-2.3cm}
\hbox{\epsfxsize=11.0cm\hspace{2.0cm}\epsfbox{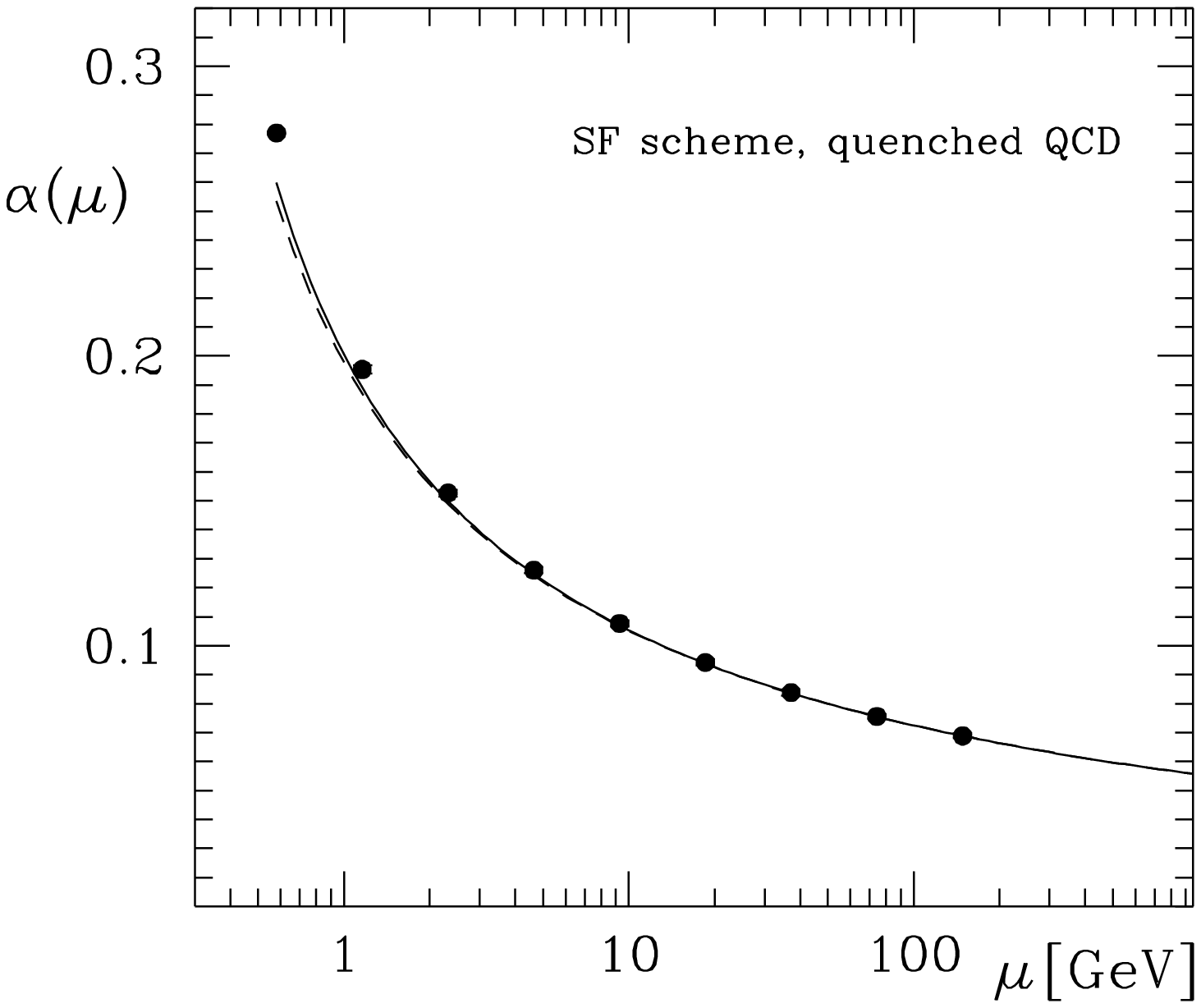}}
\vspace{-1.2cm}
\begin{center}
\footnotesize
Figure~3: Simulation results for the running coupling 
$\alpha=\gbar^2/4\pi$ in 
the SF scheme (full circles). The solid (dashed) lines are obtained
by integrating the perturbative evolution equation, starting at the 
right-most data point and using the 3-loop (2-loop) expression
for the $\beta$-function.
\end{center}
\vspace{0.0cm plus 0.3cm}
\end{figure}

Figure~3 shows the scale evolution of the running coupling
in the SF scheme, which is the particular 
finite-volume scheme that has been employed.
The data points are separated by scale factors of $2$,
i.e.~the recursion has been applied $8$ times. 
At the higher energies the scale dependence of the 
coupling is accurately reproduced by
the perturbative evolution, which has
recently been worked out to three loops in this scheme~\cite{Bode}.
The perturbative region has thus safely been reached and,
using the $3$-loop evolution in the range $\alpha\leq0.08$, 
one obtains~\cite{FSTe}
\begin{equation}
  \Lambda^{(0)}_{\smallmsbar}=251\pm21\;\MeV.
\end{equation}
The index $\raise1pt\hbox{$\scriptstyle(0)$}$ 
reminds us that this number is for quenched QCD which formally
corresponds to zero flavours of light sea quarks.
Otherwise no uncontrolled approximations have been made and 
Eq.~(14) thus is a solid result.

\begin{figure}[t]
\vspace{-2.3cm}
\hbox{\epsfxsize=11.0cm\hspace{2.0cm}\epsfbox{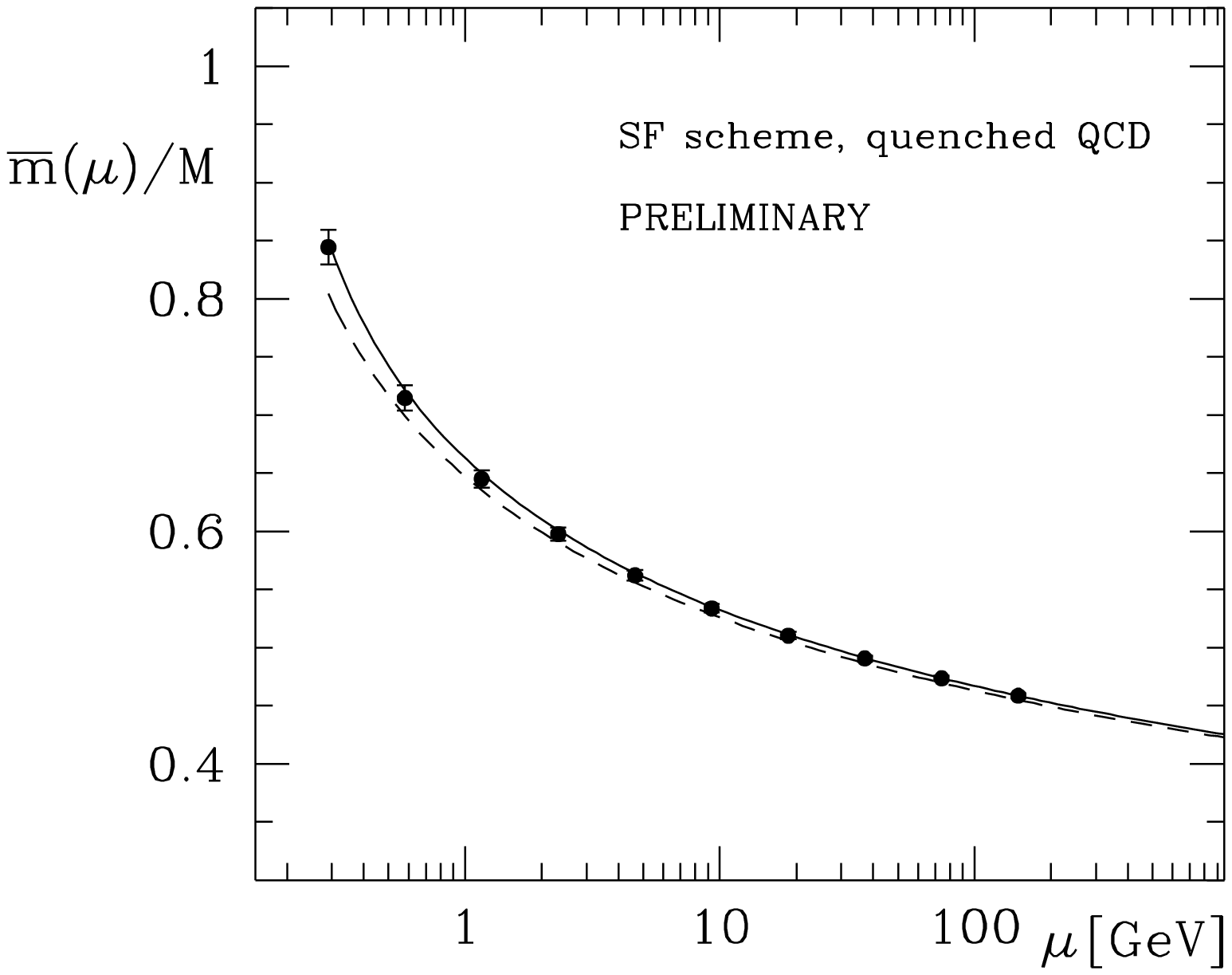}}
\vspace{-1.2cm}
\begin{center}
\footnotesize
Figure~4: Simulation results for the running quark mass in the SF scheme.
The solid (dashed) lines are obtained 
using the 2-loop (1-loop) expression for the anomalous mass dimension.
\end{center}
\vspace{0.0cm plus 0.3cm}
\end{figure}

The scale evolution of the quark masses in the SF scheme
is also accurately matched by perturbation theory
and the renormalization group invariant masses are hence
easily obtained [cf.~Eq.~(13)].
Some preliminary simulation results~\cite{FSTe} for 
the flavour-independent ratio $\mbar/M$ are plotted in Figure~4. 
The left-most data point corresponds to 
a normalization mass $\mu$ around $290$ MeV and 
a ratio $M/\mbar=1.18(2)$.
This factor provides the required link between 
the low-energy and the perturbative regime of the theory.
To complete the computation of (say) the strange 
quark mass in the $\msbar$ scheme, one still needs
to go through a few steps, but the renormalization problem has
been solved at this point and what is left to do 
are some standard calculations of meson masses and of 
the vacuum-to-kaon matrix element of 
the unrenormalized pseudo-scalar density.

The fact that the curves in Figures~3 and 4 agree so well with
the data down to very low energies
should not be given too much significance.
Rather than a general feature of the theory, the absence of 
large non-perturbative corrections to the scale evolution 
should be taken as a property of the chosen renormalization scheme.
Other schemes behave differently in this respect and there is 
usually no way to tell in advance at which energy the 
perturbative scaling sets in.

\vfill\eject

\section{Concluding remarks}

The theoretical developments described in this talk lead to 
a better understanding of the continuum limit and of the 
parameter and operator renormalization in lattice QCD.
In particular, using improved actions and the new 
techniques for non-perturbative renormali\-zation,
one will be able to obtain more reliable results 
and to approach difficult problems such as the calculation 
of moments of structure functions~\cite{SFa} and
$K\to\pi\pi$ decay rates~\cite{KPPa,KPPb} 
with greater confidence.

Most of the examples and results that have been mentioned here
refer to quenched~QCD, but the 
theoretical discussion also applies to the full theory with
any number of sea quarks.
At this point the bottleneck are the
simulation algorithms, which remain to be rather inefficient
when sea quark effects are included.
Continuous progress is however being made~\cite{Guesken}
and the new generation of dedicated computers
will no doubt allow the lattice theorists to move a big step forward
in this area too.

While there are many indications that O($a$) improvement 
and other forms of improvement are successful,
the conclusion that this will lead to dramatic savings
in computer time, ultimately allowing the solution of lattice QCD
on a PC~\cite{PCa,PCb,PCc}, is not justified.
Fast computers are indispensible
if one is interested in obtaining good control on the systematic errors. 
They are also needed to tackle the more complicated
physics issues mentioned before and the inclusion of sea quark 
effects is clearly beyond the capabilities of present-day PC's.

\section*{Acknowledgements}

I would like to thank Guido Martinelli,  
Hubert Simma, Stefan Sint, Rainer Sommer, Hartmut Wittig
and Tomoteru Yoshie for helpful correspondence
and Peter Weisz for critical comments during the preparation of this talk.

\end{document}